%% file: LATT2012_PoS.tex
\documentclass{PoS}

\graphicspath{{figs/}} 
\usepackage{mystyle} 
\usepackage{color}
\usepackage{caption, subcaption}
\let\normalcolor\relax

\newcommand{\ot}{\otimes}
\newcommand{\ct}[2]{(#1 \otimes #2)}

\title{Non-perturbative Renormalization of Improved Staggered Bilinears}

\ShortTitle{NPR of Improved Staggered Bilinears}

\author{\speaker{Andrew T. Lytle}\\ School of Physics and Astronomy, University
of Southampton, Southampton SO17 1BJ, UK\\ E-mail:
        \email{atlytle@gmail.com}\\ \hfill \it SHEP-1243}

\author{Stephen R. Sharpe\\ 
        Department of Physics, University of Washington, Seattle WA 98195-1560, USA\\
        E-mail: \email{srsharpe@uw.edu}}
\abstract{We compute $Z$-factors for general staggered bilinears on fine ($a\approx0.09\;$fm) MILC ensembles using both asqtad 
and HYP-smeared valence actions, comparing the results to the predictions of
one-loop perturbation theory.  This is an extension of previous work on the coarse ($a\approx0.12\;$fm) MILC
ensembles. It provides a laboratory for studying NPR methodology 
in the staggered context, and is an important stepping stone for fully non-perturbative
matching factors in ongoing computations of $B_K$ and other weak matrix elements.
We also implement non-exceptional {\nobreak RI/SMOM} renormalization conditions using the asqtad action and present first results.}

\FullConference{The 30th International Symposium on Lattice Field Theory\\
                 June 24 -- 29,  2012\\
                 Cairns, Australia}

\begin{document}
\section{Introduction}
Renormalization of lattice operators is an essential component for
many precision applications of lattice QCD.  Amongst these are the
determination of quark masses and weak interaction phenomenology,
both of which require matching operators to continuum
renormalization schemes.

Non-perturbative renormalization (NPR)~\cite{NPR} is a technique for
performing such matching calculations using lattice simulations
directly, and does not require the use of lattice perturbation
theory. 
Instead, it requires that the renormalization scale
$\mu$ is well separated from both the cutoff scale of the simulation,
$1/a$, and the scale at which non-perturbative QCD effects become
significant:
\begin{equation}
\Lambda_\text{QCD} \ll \mu \ll 1/a \,.
\end{equation}

There are relatively few NPR studies using staggered fermions~\cite{NPRAoki,Lytlelat09,Lytlethesis,Sharpelat11,Kim:2012ng}.  
We have calculated matching factors for all staggered fermion bilinears spread over a $2^4$ hypercube, 
determining their renormalization scale dependence and comparing them to the predictions of 
one-loop lattice perturbation theory (PT)~\cite{KLS}.
We use both HYP-smeared and asqtad valence fermions on an asqtad sea.
Previously we used the MILC coarse ensembles~\cite{Sharpelat11}; here,
in Section~\ref{sec:e}, we extend the results to the fine lattices.

The bilinear operators form the building blocks for the four-quark operators used
for weak interaction phenomenology.  An ongoing computation of $B_K$ using HYP-smeared
staggered fermions finds $\hat{B}_K= 0.727\pm0.004(\text{stat})\pm0.038(\text{sys})$~\cite{B_K}.   
The dominant error comes from using the one-loop perturbative matching factor, and
is estimated to be 4.4\% assuming that two-loop effects are of size $\alpha^2$.
The present work is an important first step towards fully non-perturbative matching,
which will both reduce the error and improve its reliability.
The current study will give a general indication of how accurate one-loop predictions are, and
thus indicate the reliability of the error estimate used in~\cite{B_K}.
The bilinears are also of phenomenological interest in their own right for the determination of quark masses.

The asqtad light-quark matching factor $Z_m$ was calculated non-perturbatively in~\cite{Lytlelat09}, using the RI/MOM scheme.
It is however advantageous to use a matching scheme that
is less sensitive to the effects of low-energy QCD.
In Section~\ref{sec:ne} we present results for the scalar and pseudoscalar bilinear channels using the RI/SMOM scheme.  We find a significant reduction in non-perturbative splitting between these channels and weaker mass dependence relative to the RI/MOM case.

\section{Exceptional NPR} \label{sec:e}
We compute bilinear $Z$-factors for all spin ($\gamma_\text{S}$) and taste ($\xi_\text{F}$) structures.
These
fall into 35 irreducible representations (irreps) under the lattice symmetry group, which are enumerated in Table~\ref{tab:bilins}. 
Unlike with Wilson or domain wall fermions, staggered bilinears are slightly non-local, spread over a $2^4$ hypercube and requiring gauge links in their definition. 
Their connection with the continuum bilinears may be expressed as follows:
\begin{equation} \label{bilins}
\overline{Q}(x)(\gamma_S\otimes\xi_F)Q'(x)
\cong
Z_{\gamma_{S} \otimes \xi_{F}}\sum_{A,B}
\overline{\chi}_A(n)\;
\overline{(\gamma_S\otimes\xi_F)}_{AB} 
U_{n+A,n+B}\; \chi_{B}'(n)
\,.
\end{equation}
Here $\chi_{A}(n)$ is a single-component field located at site $n+A$, where $n$ labels the hypercube and $A$ a site within the hypercube.
The $Q(x)$ are four-component Dirac spinors that also come in four ``tastes'' (indices left implicit).
$U_{n+A,n+B}$ is an average over all 
minimal length gauge paths connecting $n+A$ to $n+B$.

\begin{table}[t!]
\begin{center}
\begin{tabular}{c | l | l | l}
\hash{} links & S & V & T \\ \hline 
4 
&$\ct{\mathbf{1}}{\xi_{5}}$ 
&$\ct{\g_{\mu}}{\xi_{\mu} \xi_{5}}$ 
&\,\,$\ct{\g_{\mu} \g_{\nu}}{\xi_{\mu} \xi_{\nu} \xi_{5}}$ 
\\ 
3 
&$(\mathbf{1} \otimes \xi_{\mu} \xi_{5})$ &
 $(\g_{\mu} \ot \xi_{5}) \quad\, \ct{\g_{\mu}}{\xi_{\nu} \xi_{\rho}}$ &
 $\[\ct{\g_{\mu} \g_{\nu}}{\xi_{\mu} \xi_{5}} 
  \quad  \ct{\g_{\mu} \g_{\nu}}{\xi_{\rho}}\]$ 
\\ 
2
&$\ct{\mathbf{1}}{\xi_{\mu} \xi_{\nu}}$ 
&$(\g_{\mu} \ot \xi_{\nu}) \quad (\g_{\mu} \ot \xi_{\nu} \xi_{5})$ 
&$\[\textcolor{blue}{\ct{\g_{\mu} \g_{\nu}}{\mathbf{1}}}
  \quad \quad \,\,\, (\g_{\mu} \g_{\nu} \ot \xi_{5})\] 
  \quad (\g_{\mu} \g_{\nu} \ot \xi_{\nu} \xi_{\rho})$ 
\\ 
1
&$\ct{\mathbf{1}}{ \xi_{\mu}}$ 
&$\textcolor{blue}{\ct{\g_{\mu}}{\mathbf{1}}}
  \quad\,\,\,\, (\g_{\mu} \ot \xi_{\mu}\xi_{\nu})$ 
& $\[(\g_{\mu} \g_{\nu} \ot \xi_{\nu}) 
   \quad  (\g_{\mu} \g_{\nu} \ot \xi_{\rho} \xi_{5})\]$ 
\\
0
&$\textcolor{blue}{{\ct{\mathbf{1}}{\mathbf{1}}}}$
&$\ct{\g_{\mu}}{\xi_{\mu}}$
&\,\,$\ct{\g_{\mu} \g_{\nu}}{\xi_{\mu} \xi_{\nu}}$
\end{tabular}
\end{center}
\caption{Covariant bilinears forming irreps of the lattice symmetry
group. 
Indices $\mu$, $\nu$ and $\rho$ are summed from $1-4$, except
that all are different. Pseudoscalar and axial bilinears are not listed:
they can be obtained from scalar and vector, respectively, by
multiplication by $\gamma_5\otimes\xi_5$. Bilinears related
in this way have the same matching factors. This operation also implies
the identity of the Z-factors for the tensor bilinears
within square brackets.
Bilinears marked in \textcolor{blue}{blue} are 
used as the denominators of the ratios discussed in the text.}
\label{tab:bilins}
\end{table}

We compute quark propagators on coarse ($a \approx 0.12\;$fm) and fine ($a \approx 0.09\;$fm) MILC gauge ensembles, using both asqtad and HYP-smeared valence actions at three 
masses for which $am_{\text{val}} = am_{\text{sea}}$.
The use of momentum sources results in small statistical errors with relatively few (${\sim 10}$) configurations.  One feature that distinguishes NPR using staggered fermions is that for each physical momentum $p$, 16 inversions must be performed corresponding to $p + \pi A$
and the resulting momentum-space propagator is a $16 \times 16$ matrix.

The gauge links used in the the construction of the bilinears are smeared.  For the HYP valence action we use HYP smearing; for the asqtad valence action we use ``Fat 7 + Lepage'' smearing.

We impose RI'/MOM renormalization conditions on the bilinears of Equation~\eqref{bilins}
for a number of physical momenta ($\sim 10$) at each mass, and extrapolate these results
to the chiral limit. The pseudoscalar channel has a non-perturbative $1/m$ dependence
and we do not present the results for it here.

We present our comparisons to PT in terms of ratios of $Z$-factors with the
same spin-structure but different tastes.  This makes the perturbative predictions especially
simple because the continuum running cancels in the ratio:
\begin{equation}
\frac{Z_{S\otimes F1}(p)}{Z_{S\otimes F2}(p)} = 1 + \frac{\alpha(\mu_0)}{4\pi} 
\left[
C^{\rm LAT}_{S\otimes F2} - C^{\rm LAT}_{S\otimes F1}
\right] \,,
\end{equation}
where $\mu_0$ is on the order of the lattice scale and is not related to the renormalization scale $p$.
Since the ratios calculated non-perturbatively  are sensitive to lattice artifacts and low-energy effects of QCD, their variation with respect to $p$ provides an indicator as to whether one is in the NPR window.

\begin{figure}[t]
\begin{center}
\input{./figs/HYP_fine_ratios_both}
\caption{Comparison of Z-factor ratios
for vector, axial, tensor and scalar bilinears to perturbation theory
for HYP fermions on fine MILC lattices.  Horizontal lines show perturbative predictions, with
solid/dotted lines showing results with/without mean-field improvement.
Results are in the chiral limit for the momentum described in the
text. 
}
\label{fig:HYP_ratios}
\end{center}
\end{figure}
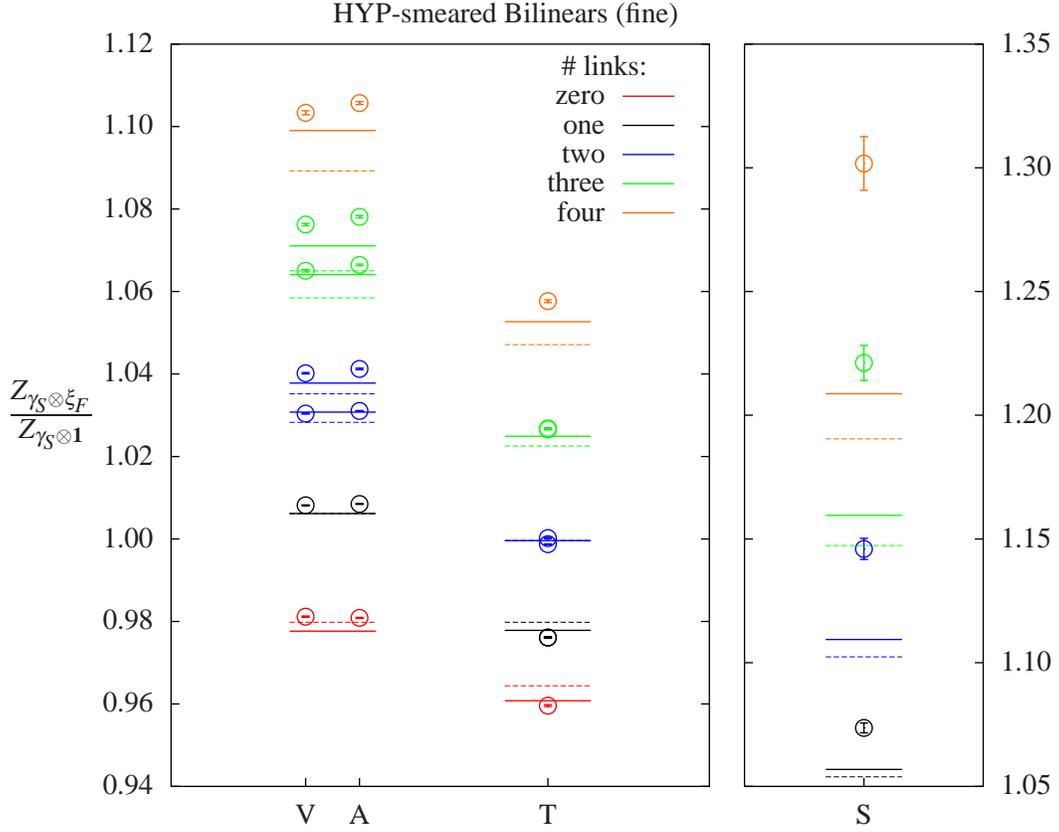

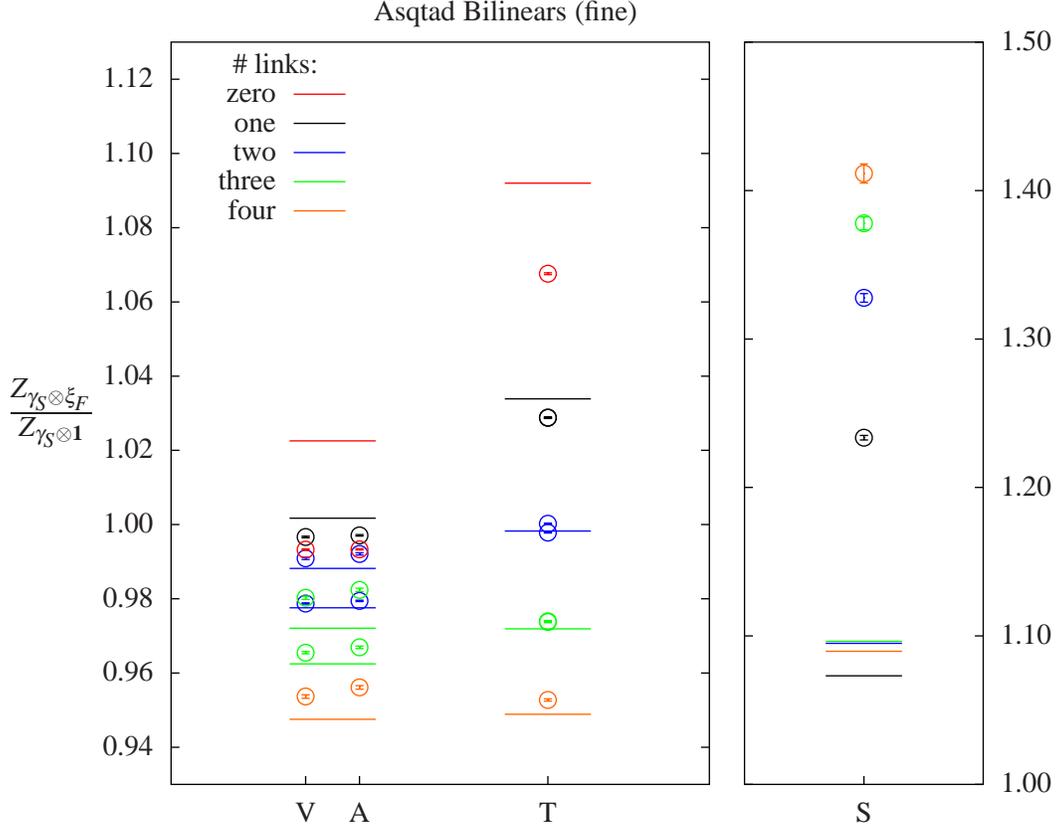
\begin{figure}[h]
\begin{center}
\input{./figs/ASQ_fine_ratios_both}
\caption{As for Fig.~\protect\ref{fig:HYP_ratios} except for asqtad fermions.
Only results from mean-field improved PT are shown.
}\label{fig:ASQ_ratios}
\end{center}
\end{figure}

\begin{figure}
\centering
\begin{subfigure}[h]{0.49\textwidth}
\input{./figs/HYP_fine_ratios_vectors}
\end{subfigure}
\begin{subfigure}[h]{0.49\textwidth}
\input{./figs/ASQ_fine_ratios_vectors}
\end{subfigure}

\begin{subfigure}[h]{0.49\textwidth}
\input{./figs/HYP_fine_ratios_tensors}
\end{subfigure}
\begin{subfigure}[h]{0.49\textwidth}
\input{./figs/ASQ_fine_ratios_tensors}
\end{subfigure}

\begin{subfigure}[h]{0.49\textwidth}
\input{./figs/HYP_fine_ratios_scalars}
\end{subfigure}
\begin{subfigure}[h]{0.49\textwidth}
\input{./figs/ASQ_fine_ratios_scalars}
\end{subfigure}
\caption{Momentum dependence of $Z$-ratios for HYP (left column) and asqtad (right column) fermions, calculated on fine MILC lattices. The colors correspond with those of the data plotted at a single scale in Figure~\protect\ref{fig:HYP_ratios} (HYP) and Figure~\protect\ref{fig:ASQ_ratios} (asqtad).}
\label{fig:scale_dependence}
\end{figure}
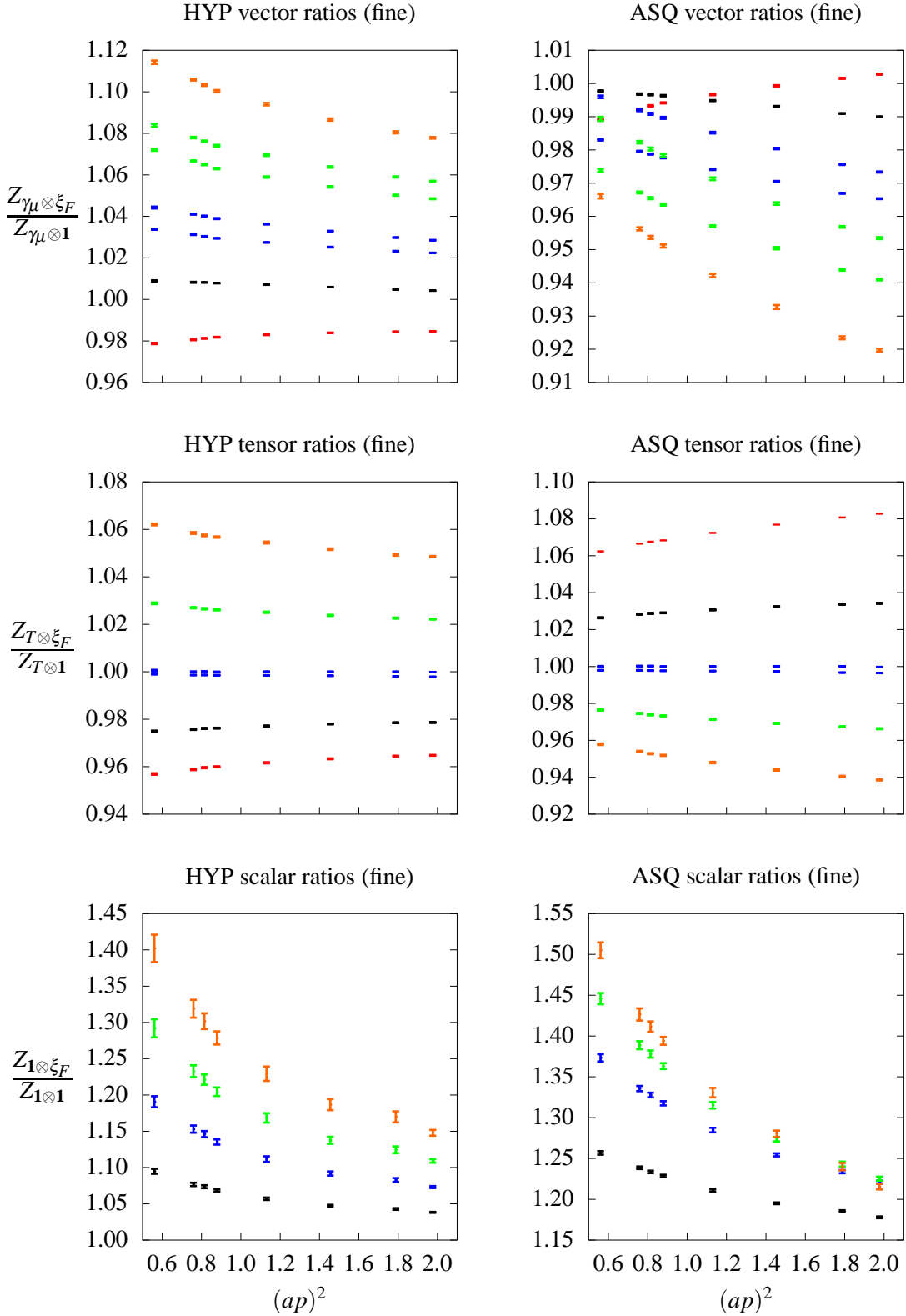

Results for the fine HYP and asqtad ratios at a single scale $p = 2.1 \text{ GeV}$  ($(ap)^2 = 0.81$) are presented in Figure~\ref{fig:HYP_ratios} and Figure~\ref{fig:ASQ_ratios}.
The perturbative results are evaluated with $\alpha(3 \text{ GeV}) = 0.24$.
For the HYP ratios, we see  good agreement with PT for the vector, axial vector, and tensor 
ratios at around the percent level.  The scalar channel shows agreement at the 5 -- 10 \%
level.  Similar results were found on the coarse MILC lattices~\cite{Sharpelat11}.
Overall, they indicate agreement at or better than the naive expectation of $\O(\alpha^2)$
truncation errors. The variation of these quantities with renormalization scale is shown in the left-hand column of Figure~\ref{fig:scale_dependence}.

For the asqtad ratios, the vector, axial vector, and tensor agree with perturbation theory at the few-percent level, 
while the scalar ratios show significant discrepancy ($\sim 25 \%$).
The variation of these quantities with renormalization scale is shown in the right-hand
column of Figure~\ref{fig:scale_dependence}.  For both the HYP and asqtad cases,
the nonperturbative values in the scalar channel more nearly approach the perturbative predictions as 
$(ap)^2$ increases.

\section{Non-exceptional NPR} \label{sec:ne}
Here we present our first results for staggered NPR using a non-exceptional scheme~(RI/SMOM)~\cite{SMOM}.
Non-exceptional schemes have reduced sensitivity to nonperturbative effects compared to their
exceptional counterparts, increasing the range of the NPR window.
They achieve this by using a kinematic setup that retains a single kinematic invariant but for which no channel has zero momentum flow 
(i.e. there are no ``exceptional'' channels).  In the case of bilinears, the flow through the external quark lines
satisfies $p_1^2 = p_2^2 = (p_1 - p_2)^2$.
Figure~\ref{fig:SminusP} shows the nonperturbative splitting between the scalar and pseudoscalar channels
in the exceptional and non-exceptional cases.
Figure~\ref{fig:LambdaS} compares the mass dependence of the scalar channel for the exceptional
and non-exceptional cases.  The non-exceptional setup has a much weaker mass dependence.
This will improve the determination of $Z_m$ using the asqtad action, where the strange sea-quark
mass is fixed and cannot be extrapolated to zero.

\begin{figure}
\centering
\begin{subfigure}[h]{0.49\textwidth}
\input{./figs/SminusP}
\caption{Non-perturbative splitting of the scalar and pseudoscalar channels using
an exceptional (E) and\\ non-exceptional (NE) scheme.}
\label{fig:SminusP}
\end{subfigure} 
\begin{subfigure}[h]{0.49\textwidth}
\input{./figs/LambdaS}
\caption{Mass dependence of the scalar channel in the \\exceptional (E) and non-exceptional (NE)
schemes. \\}
\label{fig:LambdaS}
\end{subfigure}
\end{figure}

\section{Conclusion}
We have presented non-perturbative matching factors for staggered bilinears using both
HYP-smeared and asqtad valence actions, comparing our results to the predictions of
one-loop perturbation theory over a range of momenta.
We generally find good agreement (up to $\O(\alpha^2)$) for the $Z$-ratios presented here,
except in the asqtad scalar channel (particularly for small $(ap)^2$).

We also present the first study of non-exceptional schemes using staggered fermions.
We verify that nonperturbative splitting between the scalar and pseudoscalar channels
is greatly reduced relative to the exceptional case. 
The mass dependence of the scalar channel is also much milder, which should lead to 
an improvement in the determination of $Z_m$ using this action.

Detailed descriptions of these results and associated theoretical work
are in preparation~\cite{inprep}.

\section{Acknowledgements}
The work of A.~Lytle is supported by STFC grant ST/J000396/1.
S.~Sharpe is supported in part by the US DOE grant
no.~DE-FG02-96ER40956.
Computations were carried out on USQCD Collaboration 
clusters at Fermilab.
The USQCD Collaboration is
funded by the Office of Science of the U.S. Department of Energy.

\end{document}

%% file: figs/HYP_fine_ratios_both.tex
\begingroup
  \makeatletter
  \providecommand\color[2][]{%
    \GenericError{(gnuplot) \space\space\space\@spaces}{%
      Package color not loaded in conjunction with
      terminal option `colourtext'%
    }{See the gnuplot documentation for explanation.%
    }{Either use 'blacktext' in gnuplot or load the package
      color.sty in LaTeX.}%
    \renewcommand\color[2][]{}%
  }%
  \providecommand\includegraphics[2][]{%
    \GenericError{(gnuplot) \space\space\space\@spaces}{%
      Package graphicx or graphics not loaded%
    }{See the gnuplot documentation for explanation.%
    }{The gnuplot epslatex terminal needs graphicx.sty or graphics.sty.}%
    \renewcommand\includegraphics[2][]{}%
  }%
  \providecommand\rotatebox[2]{#2}%
  \@ifundefined{ifGPcolor}{%
    \newif\ifGPcolor
    \GPcolortrue
  }{}%
  \@ifundefined{ifGPblacktext}{%
    \newif\ifGPblacktext
    \GPblacktexttrue
  }{}%
  \let\gplgaddtomacro\g@addto@macro
  \gdef\gplbacktext{}%
  \gdef\gplfronttext{}%
  \makeatother
  \ifGPblacktext
    \def\colorrgb#1{}%
    \def\colorgray#1{}%
  \else
    \ifGPcolor
      \def\colorrgb#1{\color[rgb]{#1}}%
      \def\colorgray#1{\color[gray]{#1}}%
      \expandafter\def\csname LTw\endcsname{\color{white}}%
      \expandafter\def\csname LTb\endcsname{\color{black}}%
      \expandafter\def\csname LTa\endcsname{\color{black}}%
      \expandafter\def\csname LT0\endcsname{\color[rgb]{1,0,0}}%
      \expandafter\def\csname LT1\endcsname{\color[rgb]{0,1,0}}%
      \expandafter\def\csname LT2\endcsname{\color[rgb]{0,0,1}}%
      \expandafter\def\csname LT3\endcsname{\color[rgb]{1,0,1}}%
      \expandafter\def\csname LT4\endcsname{\color[rgb]{0,1,1}}%
      \expandafter\def\csname LT5\endcsname{\color[rgb]{1,1,0}}%
      \expandafter\def\csname LT6\endcsname{\color[rgb]{0,0,0}}%
      \expandafter\def\csname LT7\endcsname{\color[rgb]{1,0.3,0}}%
      \expandafter\def\csname LT8\endcsname{\color[rgb]{0.5,0.5,0.5}}%
    \else
      \def\colorrgb#1{\color{black}}%
      \def\colorgray#1{\color[gray]{#1}}%
      \expandafter\def\csname LTw\endcsname{\color{white}}%
      \expandafter\def\csname LTb\endcsname{\color{black}}%
      \expandafter\def\csname LTa\endcsname{\color{black}}%
      \expandafter\def\csname LT0\endcsname{\color{black}}%
      \expandafter\def\csname LT1\endcsname{\color{black}}%
      \expandafter\def\csname LT2\endcsname{\color{black}}%
      \expandafter\def\csname LT3\endcsname{\color{black}}%
      \expandafter\def\csname LT4\endcsname{\color{black}}%
      \expandafter\def\csname LT5\endcsname{\color{black}}%
      \expandafter\def\csname LT6\endcsname{\color{black}}%
      \expandafter\def\csname LT7\endcsname{\color{black}}%
      \expandafter\def\csname LT8\endcsname{\color{black}}%
    \fi
  \fi
  \setlength{\unitlength}{0.0500bp}%
  \begin{picture}(7200.00,6480.00)%
      \csname LTb\endcsname%
      \put(3600,6260){\makebox(0,0){\strut{}HYP-smeared Bilinears (fine)}}%
    \gplgaddtomacro\gplbacktext{%
      \put(946,440){\makebox(0,0)[r]{\strut{}0.94}}%
      \put(946,1062){\makebox(0,0)[r]{\strut{}0.96}}%
      \put(946,1684){\makebox(0,0)[r]{\strut{}0.98}}%
      \put(946,2306){\makebox(0,0)[r]{\strut{}1.00}}%
      \put(946,2928){\makebox(0,0)[r]{\strut{}1.02}}%
      \put(946,3551){\makebox(0,0)[r]{\strut{}1.04}}%
      \put(946,4173){\makebox(0,0)[r]{\strut{}1.06}}%
      \put(946,4795){\makebox(0,0)[r]{\strut{}1.08}}%
      \put(946,5417){\makebox(0,0)[r]{\strut{}1.10}}%
      \put(946,6039){\makebox(0,0)[r]{\strut{}1.12}}%
      \put(2092,220){\makebox(0,0){\strut{}V}}%
      \put(2498,220){\makebox(0,0){\strut{}A}}%
      \put(3918,220){\makebox(0,0){\strut{}T}}%
      \put(176,3239){\makebox(0,0){\Large $\frac{Z_{\gamma_S \otimes \xi_F}}{Z_{\gamma_S \otimes \mathbf{1}}}$}}%
    }%
    \gplgaddtomacro\gplfronttext{%
      \put(4346,5866){\makebox(0,0){\strut{}$\hash$ links:}}%
      \csname LTb\endcsname%
      \put(4349,5646){\makebox(0,0)[r]{\strut{}zero}}%
      \csname LTb\endcsname%
      \put(4349,5426){\makebox(0,0)[r]{\strut{}one}}%
      \csname LTb\endcsname%
      \put(4349,5206){\makebox(0,0)[r]{\strut{}two}}%
      \csname LTb\endcsname%
      \put(4349,4986){\makebox(0,0)[r]{\strut{}three}}%
      \csname LTb\endcsname%
      \put(4349,4766){\makebox(0,0)[r]{\strut{}four}}%
    }%
    \gplgaddtomacro\gplbacktext{%
      \csname LTb\endcsname%
      \put(6300,220){\makebox(0,0){\strut{}S}}%
      \put(7331,440){\makebox(0,0)[l]{\strut{}1.05}}%
      \put(7331,1373){\makebox(0,0)[l]{\strut{}1.10}}%
      \put(7331,2306){\makebox(0,0)[l]{\strut{}1.15}}%
      \put(7331,3240){\makebox(0,0)[l]{\strut{}1.20}}%
      \put(7331,4173){\makebox(0,0)[l]{\strut{}1.25}}%
      \put(7331,5106){\makebox(0,0)[l]{\strut{}1.30}}%
      \put(7331,6039){\makebox(0,0)[l]{\strut{}1.35}}%
    }%
    \gplgaddtomacro\gplfronttext{%
    }%
    \gplbacktext
    \put(0,0){\includegraphics{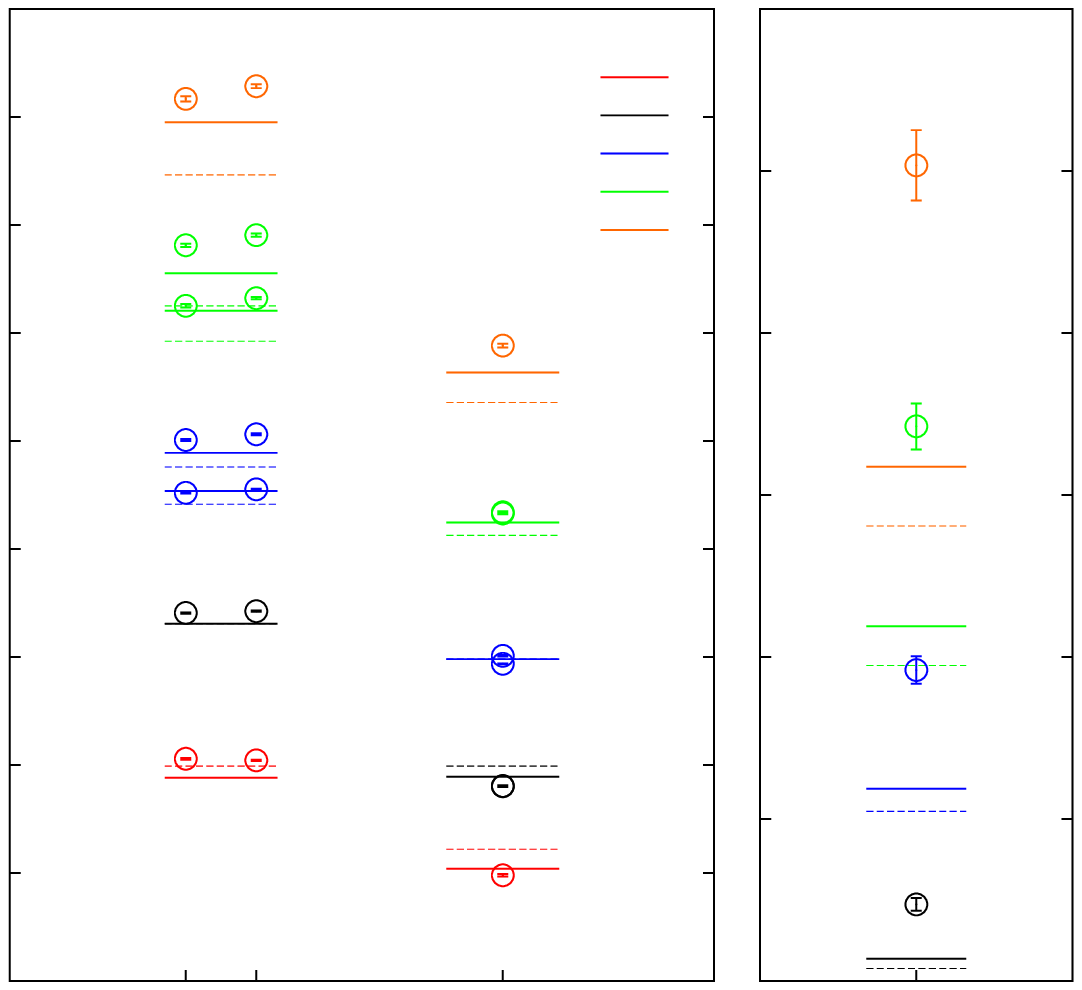}}%
    \gplfronttext
  \end{picture}%
\endgroup

%% file: figs/ASQ_fine_ratios_both.tex
\begingroup
  \makeatletter
  \providecommand\color[2][]{%
    \GenericError{(gnuplot) \space\space\space\@spaces}{%
      Package color not loaded in conjunction with
      terminal option `colourtext'%
    }{See the gnuplot documentation for explanation.%
    }{Either use 'blacktext' in gnuplot or load the package
      color.sty in LaTeX.}%
    \renewcommand\color[2][]{}%
  }%
  \providecommand\includegraphics[2][]{%
    \GenericError{(gnuplot) \space\space\space\@spaces}{%
      Package graphicx or graphics not loaded%
    }{See the gnuplot documentation for explanation.%
    }{The gnuplot epslatex terminal needs graphicx.sty or graphics.sty.}%
    \renewcommand\includegraphics[2][]{}%
  }%
  \providecommand\rotatebox[2]{#2}%
  \@ifundefined{ifGPcolor}{%
    \newif\ifGPcolor
    \GPcolortrue
  }{}%
  \@ifundefined{ifGPblacktext}{%
    \newif\ifGPblacktext
    \GPblacktexttrue
  }{}%
  \let\gplgaddtomacro\g@addto@macro
  \gdef\gplbacktext{}%
  \gdef\gplfronttext{}%
  \makeatother
  \ifGPblacktext
    \def\colorrgb#1{}%
    \def\colorgray#1{}%
  \else
    \ifGPcolor
      \def\colorrgb#1{\color[rgb]{#1}}%
      \def\colorgray#1{\color[gray]{#1}}%
      \expandafter\def\csname LTw\endcsname{\color{white}}%
      \expandafter\def\csname LTb\endcsname{\color{black}}%
      \expandafter\def\csname LTa\endcsname{\color{black}}%
      \expandafter\def\csname LT0\endcsname{\color[rgb]{1,0,0}}%
      \expandafter\def\csname LT1\endcsname{\color[rgb]{0,1,0}}%
      \expandafter\def\csname LT2\endcsname{\color[rgb]{0,0,1}}%
      \expandafter\def\csname LT3\endcsname{\color[rgb]{1,0,1}}%
      \expandafter\def\csname LT4\endcsname{\color[rgb]{0,1,1}}%
      \expandafter\def\csname LT5\endcsname{\color[rgb]{1,1,0}}%
      \expandafter\def\csname LT6\endcsname{\color[rgb]{0,0,0}}%
      \expandafter\def\csname LT7\endcsname{\color[rgb]{1,0.3,0}}%
      \expandafter\def\csname LT8\endcsname{\color[rgb]{0.5,0.5,0.5}}%
    \else
      \def\colorrgb#1{\color{black}}%
      \def\colorgray#1{\color[gray]{#1}}%
      \expandafter\def\csname LTw\endcsname{\color{white}}%
      \expandafter\def\csname LTb\endcsname{\color{black}}%
      \expandafter\def\csname LTa\endcsname{\color{black}}%
      \expandafter\def\csname LT0\endcsname{\color{black}}%
      \expandafter\def\csname LT1\endcsname{\color{black}}%
      \expandafter\def\csname LT2\endcsname{\color{black}}%
      \expandafter\def\csname LT3\endcsname{\color{black}}%
      \expandafter\def\csname LT4\endcsname{\color{black}}%
      \expandafter\def\csname LT5\endcsname{\color{black}}%
      \expandafter\def\csname LT6\endcsname{\color{black}}%
      \expandafter\def\csname LT7\endcsname{\color{black}}%
      \expandafter\def\csname LT8\endcsname{\color{black}}%
    \fi
  \fi
  \setlength{\unitlength}{0.0500bp}%
  \begin{picture}(7200.00,6480.00)%
      \csname LTb\endcsname%
      \put(3600,6260){\makebox(0,0){\strut{}Asqtad Bilinears (fine)}}%
    \gplgaddtomacro\gplbacktext{%
      \put(946,720){\makebox(0,0)[r]{\strut{}0.94}}%
      \put(946,1280){\makebox(0,0)[r]{\strut{}0.96}}%
      \put(946,1840){\makebox(0,0)[r]{\strut{}0.98}}%
      \put(946,2400){\makebox(0,0)[r]{\strut{}1.00}}%
      \put(946,2960){\makebox(0,0)[r]{\strut{}1.02}}%
      \put(946,3519){\makebox(0,0)[r]{\strut{}1.04}}%
      \put(946,4079){\makebox(0,0)[r]{\strut{}1.06}}%
      \put(946,4639){\makebox(0,0)[r]{\strut{}1.08}}%
      \put(946,5199){\makebox(0,0)[r]{\strut{}1.10}}%
      \put(946,5759){\makebox(0,0)[r]{\strut{}1.12}}%
      \put(2092,220){\makebox(0,0){\strut{}V}}%
      \put(2498,220){\makebox(0,0){\strut{}A}}%
      \put(3918,220){\makebox(0,0){\strut{}T}}%
      \put(176,3239){\makebox(0,0){\Large $\frac{Z_{\gamma_S \otimes \xi_F}}{Z_{\gamma_S \otimes \mathbf{1}}}$}}%
    }%
    \gplgaddtomacro\gplfronttext{%
      \put(1867,5866){\makebox(0,0){\strut{}$\hash$ links:}}%
      \csname LTb\endcsname%
      \put(1870,5646){\makebox(0,0)[r]{\strut{}zero}}%
      \csname LTb\endcsname%
      \put(1870,5426){\makebox(0,0)[r]{\strut{}one}}%
      \csname LTb\endcsname%
      \put(1870,5206){\makebox(0,0)[r]{\strut{}two}}%
      \csname LTb\endcsname%
      \put(1870,4986){\makebox(0,0)[r]{\strut{}three}}%
      \csname LTb\endcsname%
      \put(1870,4766){\makebox(0,0)[r]{\strut{}four}}%
    }%
    \gplgaddtomacro\gplbacktext{%
      \csname LTb\endcsname%
      \put(6300,220){\makebox(0,0){\strut{}S}}%
      \put(7331,440){\makebox(0,0)[l]{\strut{}1.00}}%
      \put(7331,1560){\makebox(0,0)[l]{\strut{}1.10}}%
      \put(7331,2680){\makebox(0,0)[l]{\strut{}1.20}}%
      \put(7331,3799){\makebox(0,0)[l]{\strut{}1.30}}%
      \put(7331,4919){\makebox(0,0)[l]{\strut{}1.40}}%
      \put(7331,6039){\makebox(0,0)[l]{\strut{}1.50}}%
    }%
    \gplgaddtomacro\gplfronttext{%
    }%
    \gplbacktext
    \put(0,0){\includegraphics{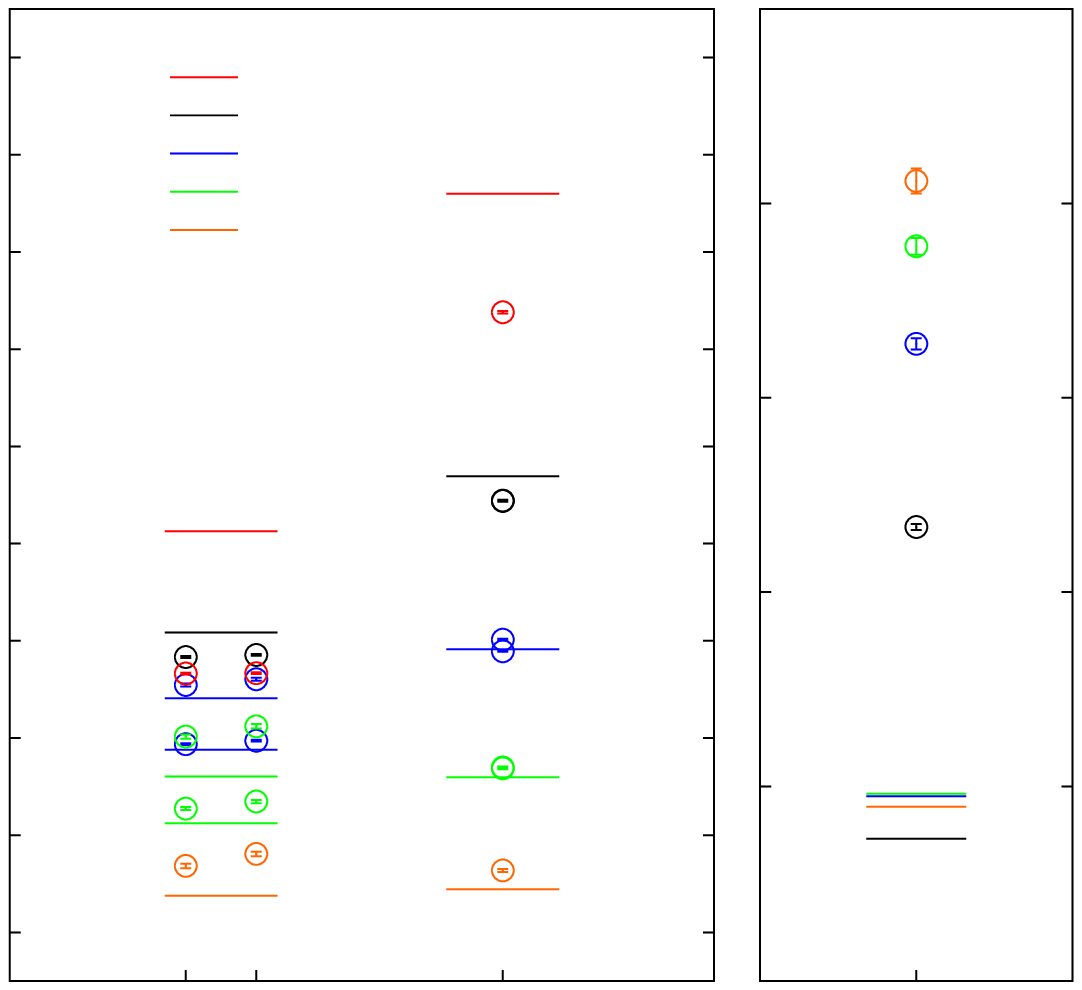}}%
    \gplfronttext
  \end{picture}%
\endgroup

%% file: figs/HYP_fine_ratios_vectors.tex
\begingroup
  \makeatletter
  \providecommand\color[2][]{%
    \GenericError{(gnuplot) \space\space\space\@spaces}{%
      Package color not loaded in conjunction with
      terminal option `colourtext'%
    }{See the gnuplot documentation for explanation.%
    }{Either use 'blacktext' in gnuplot or load the package
      color.sty in LaTeX.}%
    \renewcommand\color[2][]{}%
  }%
  \providecommand\includegraphics[2][]{%
    \GenericError{(gnuplot) \space\space\space\@spaces}{%
      Package graphicx or graphics not loaded%
    }{See the gnuplot documentation for explanation.%
    }{The gnuplot epslatex terminal needs graphicx.sty or graphics.sty.}%
    \renewcommand\includegraphics[2][]{}%
  }%
  \providecommand\rotatebox[2]{#2}%
  \@ifundefined{ifGPcolor}{%
    \newif\ifGPcolor
    \GPcolortrue
  }{}%
  \@ifundefined{ifGPblacktext}{%
    \newif\ifGPblacktext
    \GPblacktexttrue
  }{}%
  \let\gplgaddtomacro\g@addto@macro
  \gdef\gplbacktext{}%
  \gdef\gplfronttext{}%
  \makeatother
  \ifGPblacktext
    \def\colorrgb#1{}%
    \def\colorgray#1{}%
  \else
    \ifGPcolor
      \def\colorrgb#1{\color[rgb]{#1}}%
      \def\colorgray#1{\color[gray]{#1}}%
      \expandafter\def\csname LTw\endcsname{\color{white}}%
      \expandafter\def\csname LTb\endcsname{\color{black}}%
      \expandafter\def\csname LTa\endcsname{\color{black}}%
      \expandafter\def\csname LT0\endcsname{\color[rgb]{1,0,0}}%
      \expandafter\def\csname LT1\endcsname{\color[rgb]{0,1,0}}%
      \expandafter\def\csname LT2\endcsname{\color[rgb]{0,0,1}}%
      \expandafter\def\csname LT3\endcsname{\color[rgb]{1,0,1}}%
      \expandafter\def\csname LT4\endcsname{\color[rgb]{0,1,1}}%
      \expandafter\def\csname LT5\endcsname{\color[rgb]{1,1,0}}%
      \expandafter\def\csname LT6\endcsname{\color[rgb]{0,0,0}}%
      \expandafter\def\csname LT7\endcsname{\color[rgb]{1,0.3,0}}%
      \expandafter\def\csname LT8\endcsname{\color[rgb]{0.5,0.5,0.5}}%
    \else
      \def\colorrgb#1{\color{black}}%
      \def\colorgray#1{\color[gray]{#1}}%
      \expandafter\def\csname LTw\endcsname{\color{white}}%
      \expandafter\def\csname LTb\endcsname{\color{black}}%
      \expandafter\def\csname LTa\endcsname{\color{black}}%
      \expandafter\def\csname LT0\endcsname{\color{black}}%
      \expandafter\def\csname LT1\endcsname{\color{black}}%
      \expandafter\def\csname LT2\endcsname{\color{black}}%
      \expandafter\def\csname LT3\endcsname{\color{black}}%
      \expandafter\def\csname LT4\endcsname{\color{black}}%
      \expandafter\def\csname LT5\endcsname{\color{black}}%
      \expandafter\def\csname LT6\endcsname{\color{black}}%
      \expandafter\def\csname LT7\endcsname{\color{black}}%
      \expandafter\def\csname LT8\endcsname{\color{black}}%
    \fi
  \fi
  \setlength{\unitlength}{0.0500bp}%
  \begin{picture}(4320.00,3888.00)%
    \gplgaddtomacro\gplbacktext{%
      \colorrgb{0.00,0.00,0.00}%
      \put(946,220){\makebox(0,0)[r]{\strut{}0.96}}%
      \colorrgb{0.00,0.00,0.00}%
      \put(946,596){\makebox(0,0)[r]{\strut{}0.98}}%
      \colorrgb{0.00,0.00,0.00}%
      \put(946,972){\makebox(0,0)[r]{\strut{}1.00}}%
      \colorrgb{0.00,0.00,0.00}%
      \put(946,1348){\makebox(0,0)[r]{\strut{}1.02}}%
      \colorrgb{0.00,0.00,0.00}%
      \put(946,1724){\makebox(0,0)[r]{\strut{}1.04}}%
      \colorrgb{0.00,0.00,0.00}%
      \put(946,2099){\makebox(0,0)[r]{\strut{}1.06}}%
      \colorrgb{0.00,0.00,0.00}%
      \put(946,2475){\makebox(0,0)[r]{\strut{}1.08}}%
      \colorrgb{0.00,0.00,0.00}%
      \put(946,2851){\makebox(0,0)[r]{\strut{}1.10}}%
      \colorrgb{0.00,0.00,0.00}%
      \put(946,3227){\makebox(0,0)[r]{\strut{}1.12}}%
      \colorrgb{0.00,0.00,0.00}%
      \put(1256,0){\makebox(0,0){\strut{}}}%
      \colorrgb{0.00,0.00,0.00}%
      \put(1611,0){\makebox(0,0){\strut{}}}%
      \colorrgb{0.00,0.00,0.00}%
      \put(1967,0){\makebox(0,0){\strut{}}}%
      \colorrgb{0.00,0.00,0.00}%
      \put(2323,0){\makebox(0,0){\strut{}}}%
      \colorrgb{0.00,0.00,0.00}%
      \put(2678,0){\makebox(0,0){\strut{}}}%
      \colorrgb{0.00,0.00,0.00}%
      \put(3034,0){\makebox(0,0){\strut{}}}%
      \colorrgb{0.00,0.00,0.00}%
      \put(3390,0){\makebox(0,0){\strut{}}}%
      \colorrgb{0.00,0.00,0.00}%
      \put(3745,0){\makebox(0,0){\strut{}}}%
      \csname LTb\endcsname%
      \put(176,1723){\makebox(0,0){\Large $\frac{Z_{\gamma_{\mu} \otimes \xi_F}}            {Z_{\gamma_{\mu} \otimes \mathbf{1}}}$}}%
      \put(2500,-66){\makebox(0,0){\strut{}}}%
      \put(2500,3557){\makebox(0,0){\strut{}HYP vector ratios (fine)}}%
    }%
    \gplgaddtomacro\gplfronttext{%
    }%
    \gplbacktext
    \put(0,0){\includegraphics{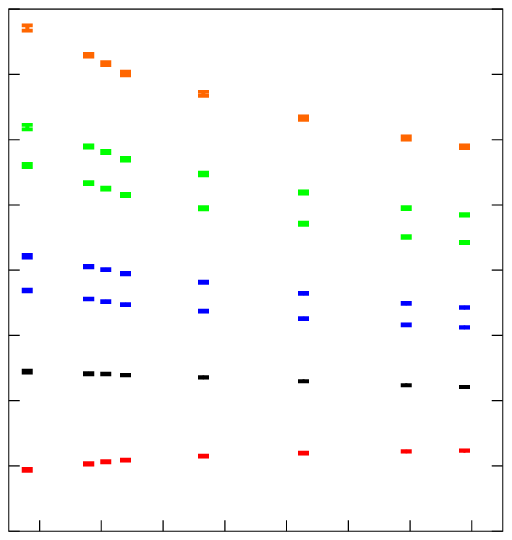}}%
    \gplfronttext
  \end{picture}%
\endgroup

%% file: figs/ASQ_fine_ratios_vectors.tex
\begingroup
  \makeatletter
  \providecommand\color[2][]{%
    \GenericError{(gnuplot) \space\space\space\@spaces}{%
      Package color not loaded in conjunction with
      terminal option `colourtext'%
    }{See the gnuplot documentation for explanation.%
    }{Either use 'blacktext' in gnuplot or load the package
      color.sty in LaTeX.}%
    \renewcommand\color[2][]{}%
  }%
  \providecommand\includegraphics[2][]{%
    \GenericError{(gnuplot) \space\space\space\@spaces}{%
      Package graphicx or graphics not loaded%
    }{See the gnuplot documentation for explanation.%
    }{The gnuplot epslatex terminal needs graphicx.sty or graphics.sty.}%
    \renewcommand\includegraphics[2][]{}%
  }%
  \providecommand\rotatebox[2]{#2}%
  \@ifundefined{ifGPcolor}{%
    \newif\ifGPcolor
    \GPcolortrue
  }{}%
  \@ifundefined{ifGPblacktext}{%
    \newif\ifGPblacktext
    \GPblacktexttrue
  }{}%
  \let\gplgaddtomacro\g@addto@macro
  \gdef\gplbacktext{}%
  \gdef\gplfronttext{}%
  \makeatother
  \ifGPblacktext
    \def\colorrgb#1{}%
    \def\colorgray#1{}%
  \else
    \ifGPcolor
      \def\colorrgb#1{\color[rgb]{#1}}%
      \def\colorgray#1{\color[gray]{#1}}%
      \expandafter\def\csname LTw\endcsname{\color{white}}%
      \expandafter\def\csname LTb\endcsname{\color{black}}%
      \expandafter\def\csname LTa\endcsname{\color{black}}%
      \expandafter\def\csname LT0\endcsname{\color[rgb]{1,0,0}}%
      \expandafter\def\csname LT1\endcsname{\color[rgb]{0,1,0}}%
      \expandafter\def\csname LT2\endcsname{\color[rgb]{0,0,1}}%
      \expandafter\def\csname LT3\endcsname{\color[rgb]{1,0,1}}%
      \expandafter\def\csname LT4\endcsname{\color[rgb]{0,1,1}}%
      \expandafter\def\csname LT5\endcsname{\color[rgb]{1,1,0}}%
      \expandafter\def\csname LT6\endcsname{\color[rgb]{0,0,0}}%
      \expandafter\def\csname LT7\endcsname{\color[rgb]{1,0.3,0}}%
      \expandafter\def\csname LT8\endcsname{\color[rgb]{0.5,0.5,0.5}}%
    \else
      \def\colorrgb#1{\color{black}}%
      \def\colorgray#1{\color[gray]{#1}}%
      \expandafter\def\csname LTw\endcsname{\color{white}}%
      \expandafter\def\csname LTb\endcsname{\color{black}}%
      \expandafter\def\csname LTa\endcsname{\color{black}}%
      \expandafter\def\csname LT0\endcsname{\color{black}}%
      \expandafter\def\csname LT1\endcsname{\color{black}}%
      \expandafter\def\csname LT2\endcsname{\color{black}}%
      \expandafter\def\csname LT3\endcsname{\color{black}}%
      \expandafter\def\csname LT4\endcsname{\color{black}}%
      \expandafter\def\csname LT5\endcsname{\color{black}}%
      \expandafter\def\csname LT6\endcsname{\color{black}}%
      \expandafter\def\csname LT7\endcsname{\color{black}}%
      \expandafter\def\csname LT8\endcsname{\color{black}}%
    \fi
  \fi
  \setlength{\unitlength}{0.0500bp}%
  \begin{picture}(4104.00,3888.00)%
    \gplgaddtomacro\gplbacktext{%
      \colorrgb{0.00,0.00,0.00}%
      \put(726,220){\makebox(0,0)[r]{\strut{}0.91}}%
      \colorrgb{0.00,0.00,0.00}%
      \put(726,521){\makebox(0,0)[r]{\strut{}0.92}}%
      \colorrgb{0.00,0.00,0.00}%
      \put(726,821){\makebox(0,0)[r]{\strut{}0.93}}%
      \colorrgb{0.00,0.00,0.00}%
      \put(726,1122){\makebox(0,0)[r]{\strut{}0.94}}%
      \colorrgb{0.00,0.00,0.00}%
      \put(726,1423){\makebox(0,0)[r]{\strut{}0.95}}%
      \colorrgb{0.00,0.00,0.00}%
      \put(726,1724){\makebox(0,0)[r]{\strut{}0.96}}%
      \colorrgb{0.00,0.00,0.00}%
      \put(726,2024){\makebox(0,0)[r]{\strut{}0.97}}%
      \colorrgb{0.00,0.00,0.00}%
      \put(726,2325){\makebox(0,0)[r]{\strut{}0.98}}%
      \colorrgb{0.00,0.00,0.00}%
      \put(726,2626){\makebox(0,0)[r]{\strut{}0.99}}%
      \colorrgb{0.00,0.00,0.00}%
      \put(726,2926){\makebox(0,0)[r]{\strut{}1.00}}%
      \colorrgb{0.00,0.00,0.00}%
      \put(726,3227){\makebox(0,0)[r]{\strut{}1.01}}%
      \colorrgb{0.00,0.00,0.00}%
      \put(1036,0){\makebox(0,0){\strut{}}}%
      \colorrgb{0.00,0.00,0.00}%
      \put(1392,0){\makebox(0,0){\strut{}}}%
      \colorrgb{0.00,0.00,0.00}%
      \put(1748,0){\makebox(0,0){\strut{}}}%
      \colorrgb{0.00,0.00,0.00}%
      \put(2104,0){\makebox(0,0){\strut{}}}%
      \colorrgb{0.00,0.00,0.00}%
      \put(2461,0){\makebox(0,0){\strut{}}}%
      \colorrgb{0.00,0.00,0.00}%
      \put(2817,0){\makebox(0,0){\strut{}}}%
      \colorrgb{0.00,0.00,0.00}%
      \put(3173,0){\makebox(0,0){\strut{}}}%
      \colorrgb{0.00,0.00,0.00}%
      \put(3529,0){\makebox(0,0){\strut{}}}%
      \csname LTb\endcsname%
      \put(176,1723){\rotatebox{-270}{\makebox(0,0){\strut{}}}}%
      \put(2282,-66){\makebox(0,0){\strut{}}}%
      \put(2282,3557){\makebox(0,0){\strut{}ASQ vector ratios (fine)}}%
    }%
    \gplgaddtomacro\gplfronttext{%
    }%
    \gplbacktext
    \put(0,0){\includegraphics{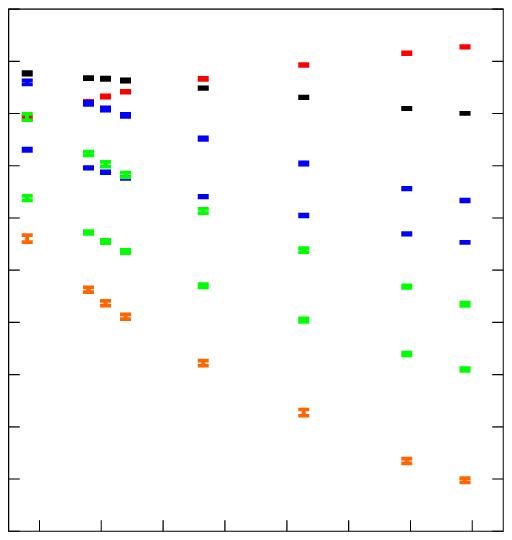}}%
    \gplfronttext
  \end{picture}%
\endgroup

%% file: figs/HYP_fine_ratios_tensors.tex
\begingroup
  \makeatletter
  \providecommand\color[2][]{%
    \GenericError{(gnuplot) \space\space\space\@spaces}{%
      Package color not loaded in conjunction with
      terminal option `colourtext'%
    }{See the gnuplot documentation for explanation.%
    }{Either use 'blacktext' in gnuplot or load the package
      color.sty in LaTeX.}%
    \renewcommand\color[2][]{}%
  }%
  \providecommand\includegraphics[2][]{%
    \GenericError{(gnuplot) \space\space\space\@spaces}{%
      Package graphicx or graphics not loaded%
    }{See the gnuplot documentation for explanation.%
    }{The gnuplot epslatex terminal needs graphicx.sty or graphics.sty.}%
    \renewcommand\includegraphics[2][]{}%
  }%
  \providecommand\rotatebox[2]{#2}%
  \@ifundefined{ifGPcolor}{%
    \newif\ifGPcolor
    \GPcolortrue
  }{}%
  \@ifundefined{ifGPblacktext}{%
    \newif\ifGPblacktext
    \GPblacktexttrue
  }{}%
  \let\gplgaddtomacro\g@addto@macro
  \gdef\gplbacktext{}%
  \gdef\gplfronttext{}%
  \makeatother
  \ifGPblacktext
    \def\colorrgb#1{}%
    \def\colorgray#1{}%
  \else
    \ifGPcolor
      \def\colorrgb#1{\color[rgb]{#1}}%
      \def\colorgray#1{\color[gray]{#1}}%
      \expandafter\def\csname LTw\endcsname{\color{white}}%
      \expandafter\def\csname LTb\endcsname{\color{black}}%
      \expandafter\def\csname LTa\endcsname{\color{black}}%
      \expandafter\def\csname LT0\endcsname{\color[rgb]{1,0,0}}%
      \expandafter\def\csname LT1\endcsname{\color[rgb]{0,1,0}}%
      \expandafter\def\csname LT2\endcsname{\color[rgb]{0,0,1}}%
      \expandafter\def\csname LT3\endcsname{\color[rgb]{1,0,1}}%
      \expandafter\def\csname LT4\endcsname{\color[rgb]{0,1,1}}%
      \expandafter\def\csname LT5\endcsname{\color[rgb]{1,1,0}}%
      \expandafter\def\csname LT6\endcsname{\color[rgb]{0,0,0}}%
      \expandafter\def\csname LT7\endcsname{\color[rgb]{1,0.3,0}}%
      \expandafter\def\csname LT8\endcsname{\color[rgb]{0.5,0.5,0.5}}%
    \else
      \def\colorrgb#1{\color{black}}%
      \def\colorgray#1{\color[gray]{#1}}%
      \expandafter\def\csname LTw\endcsname{\color{white}}%
      \expandafter\def\csname LTb\endcsname{\color{black}}%
      \expandafter\def\csname LTa\endcsname{\color{black}}%
      \expandafter\def\csname LT0\endcsname{\color{black}}%
      \expandafter\def\csname LT1\endcsname{\color{black}}%
      \expandafter\def\csname LT2\endcsname{\color{black}}%
      \expandafter\def\csname LT3\endcsname{\color{black}}%
      \expandafter\def\csname LT4\endcsname{\color{black}}%
      \expandafter\def\csname LT5\endcsname{\color{black}}%
      \expandafter\def\csname LT6\endcsname{\color{black}}%
      \expandafter\def\csname LT7\endcsname{\color{black}}%
      \expandafter\def\csname LT8\endcsname{\color{black}}%
    \fi
  \fi
  \setlength{\unitlength}{0.0500bp}%
  \begin{picture}(4320.00,3888.00)%
    \gplgaddtomacro\gplbacktext{%
      \colorrgb{0.00,0.00,0.00}%
      \put(946,220){\makebox(0,0)[r]{\strut{}0.94}}%
      \colorrgb{0.00,0.00,0.00}%
      \put(946,650){\makebox(0,0)[r]{\strut{}0.96}}%
      \colorrgb{0.00,0.00,0.00}%
      \put(946,1079){\makebox(0,0)[r]{\strut{}0.98}}%
      \colorrgb{0.00,0.00,0.00}%
      \put(946,1509){\makebox(0,0)[r]{\strut{}1.00}}%
      \colorrgb{0.00,0.00,0.00}%
      \put(946,1938){\makebox(0,0)[r]{\strut{}1.02}}%
      \colorrgb{0.00,0.00,0.00}%
      \put(946,2368){\makebox(0,0)[r]{\strut{}1.04}}%
      \colorrgb{0.00,0.00,0.00}%
      \put(946,2797){\makebox(0,0)[r]{\strut{}1.06}}%
      \colorrgb{0.00,0.00,0.00}%
      \put(946,3227){\makebox(0,0)[r]{\strut{}1.08}}%
      \colorrgb{0.00,0.00,0.00}%
      \put(1256,0){\makebox(0,0){\strut{}}}%
      \colorrgb{0.00,0.00,0.00}%
      \put(1611,0){\makebox(0,0){\strut{}}}%
      \colorrgb{0.00,0.00,0.00}%
      \put(1967,0){\makebox(0,0){\strut{}}}%
      \colorrgb{0.00,0.00,0.00}%
      \put(2323,0){\makebox(0,0){\strut{}}}%
      \colorrgb{0.00,0.00,0.00}%
      \put(2678,0){\makebox(0,0){\strut{}}}%
      \colorrgb{0.00,0.00,0.00}%
      \put(3034,0){\makebox(0,0){\strut{}}}%
      \colorrgb{0.00,0.00,0.00}%
      \put(3390,0){\makebox(0,0){\strut{}}}%
      \colorrgb{0.00,0.00,0.00}%
      \put(3745,0){\makebox(0,0){\strut{}}}%
      \csname LTb\endcsname%
      \put(176,1723){\makebox(0,0){\Large $\frac{Z_{T \otimes \xi_F}}            {Z_{T \otimes \mathbf{1}}}$}}%
      \put(2500,-66){\makebox(0,0){\strut{}}}%
      \put(2500,3557){\makebox(0,0){\strut{}HYP tensor ratios (fine)}}%
    }%
    \gplgaddtomacro\gplfronttext{%
    }%
    \gplbacktext
    \put(0,0){\includegraphics{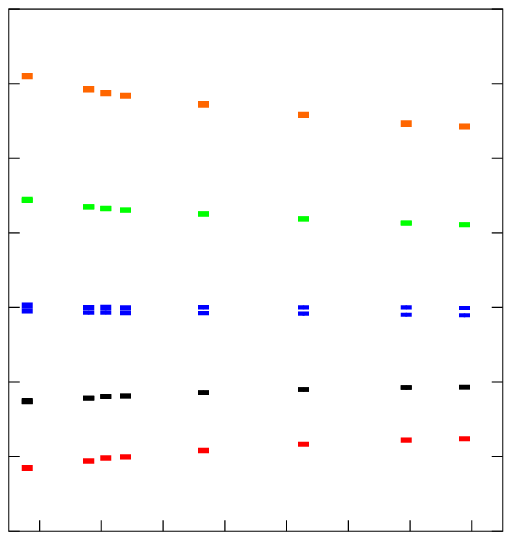}}%
    \gplfronttext
  \end{picture}%
\endgroup

%% file: figs/ASQ_fine_ratios_tensors.tex
\begingroup
  \makeatletter
  \providecommand\color[2][]{%
    \GenericError{(gnuplot) \space\space\space\@spaces}{%
      Package color not loaded in conjunction with
      terminal option `colourtext'%
    }{See the gnuplot documentation for explanation.%
    }{Either use 'blacktext' in gnuplot or load the package
      color.sty in LaTeX.}%
    \renewcommand\color[2][]{}%
  }%
  \providecommand\includegraphics[2][]{%
    \GenericError{(gnuplot) \space\space\space\@spaces}{%
      Package graphicx or graphics not loaded%
    }{See the gnuplot documentation for explanation.%
    }{The gnuplot epslatex terminal needs graphicx.sty or graphics.sty.}%
    \renewcommand\includegraphics[2][]{}%
  }%
  \providecommand\rotatebox[2]{#2}%
  \@ifundefined{ifGPcolor}{%
    \newif\ifGPcolor
    \GPcolortrue
  }{}%
  \@ifundefined{ifGPblacktext}{%
    \newif\ifGPblacktext
    \GPblacktexttrue
  }{}%
  \let\gplgaddtomacro\g@addto@macro
  \gdef\gplbacktext{}%
  \gdef\gplfronttext{}%
  \makeatother
  \ifGPblacktext
    \def\colorrgb#1{}%
    \def\colorgray#1{}%
  \else
    \ifGPcolor
      \def\colorrgb#1{\color[rgb]{#1}}%
      \def\colorgray#1{\color[gray]{#1}}%
      \expandafter\def\csname LTw\endcsname{\color{white}}%
      \expandafter\def\csname LTb\endcsname{\color{black}}%
      \expandafter\def\csname LTa\endcsname{\color{black}}%
      \expandafter\def\csname LT0\endcsname{\color[rgb]{1,0,0}}%
      \expandafter\def\csname LT1\endcsname{\color[rgb]{0,1,0}}%
      \expandafter\def\csname LT2\endcsname{\color[rgb]{0,0,1}}%
      \expandafter\def\csname LT3\endcsname{\color[rgb]{1,0,1}}%
      \expandafter\def\csname LT4\endcsname{\color[rgb]{0,1,1}}%
      \expandafter\def\csname LT5\endcsname{\color[rgb]{1,1,0}}%
      \expandafter\def\csname LT6\endcsname{\color[rgb]{0,0,0}}%
      \expandafter\def\csname LT7\endcsname{\color[rgb]{1,0.3,0}}%
      \expandafter\def\csname LT8\endcsname{\color[rgb]{0.5,0.5,0.5}}%
    \else
      \def\colorrgb#1{\color{black}}%
      \def\colorgray#1{\color[gray]{#1}}%
      \expandafter\def\csname LTw\endcsname{\color{white}}%
      \expandafter\def\csname LTb\endcsname{\color{black}}%
      \expandafter\def\csname LTa\endcsname{\color{black}}%
      \expandafter\def\csname LT0\endcsname{\color{black}}%
      \expandafter\def\csname LT1\endcsname{\color{black}}%
      \expandafter\def\csname LT2\endcsname{\color{black}}%
      \expandafter\def\csname LT3\endcsname{\color{black}}%
      \expandafter\def\csname LT4\endcsname{\color{black}}%
      \expandafter\def\csname LT5\endcsname{\color{black}}%
      \expandafter\def\csname LT6\endcsname{\color{black}}%
      \expandafter\def\csname LT7\endcsname{\color{black}}%
      \expandafter\def\csname LT8\endcsname{\color{black}}%
    \fi
  \fi
  \setlength{\unitlength}{0.0500bp}%
  \begin{picture}(4104.00,3888.00)%
    \gplgaddtomacro\gplbacktext{%
      \colorrgb{0.00,0.00,0.00}%
      \put(726,220){\makebox(0,0)[r]{\strut{}0.92}}%
      \colorrgb{0.00,0.00,0.00}%
      \put(726,554){\makebox(0,0)[r]{\strut{}0.94}}%
      \colorrgb{0.00,0.00,0.00}%
      \put(726,888){\makebox(0,0)[r]{\strut{}0.96}}%
      \colorrgb{0.00,0.00,0.00}%
      \put(726,1222){\makebox(0,0)[r]{\strut{}0.98}}%
      \colorrgb{0.00,0.00,0.00}%
      \put(726,1556){\makebox(0,0)[r]{\strut{}1.00}}%
      \colorrgb{0.00,0.00,0.00}%
      \put(726,1891){\makebox(0,0)[r]{\strut{}1.02}}%
      \colorrgb{0.00,0.00,0.00}%
      \put(726,2225){\makebox(0,0)[r]{\strut{}1.04}}%
      \colorrgb{0.00,0.00,0.00}%
      \put(726,2559){\makebox(0,0)[r]{\strut{}1.06}}%
      \colorrgb{0.00,0.00,0.00}%
      \put(726,2893){\makebox(0,0)[r]{\strut{}1.08}}%
      \colorrgb{0.00,0.00,0.00}%
      \put(726,3227){\makebox(0,0)[r]{\strut{}1.10}}%
      \colorrgb{0.00,0.00,0.00}%
      \put(1036,0){\makebox(0,0){\strut{}}}%
      \colorrgb{0.00,0.00,0.00}%
      \put(1392,0){\makebox(0,0){\strut{}}}%
      \colorrgb{0.00,0.00,0.00}%
      \put(1748,0){\makebox(0,0){\strut{}}}%
      \colorrgb{0.00,0.00,0.00}%
      \put(2104,0){\makebox(0,0){\strut{}}}%
      \colorrgb{0.00,0.00,0.00}%
      \put(2461,0){\makebox(0,0){\strut{}}}%
      \colorrgb{0.00,0.00,0.00}%
      \put(2817,0){\makebox(0,0){\strut{}}}%
      \colorrgb{0.00,0.00,0.00}%
      \put(3173,0){\makebox(0,0){\strut{}}}%
      \colorrgb{0.00,0.00,0.00}%
      \put(3529,0){\makebox(0,0){\strut{}}}%
      \csname LTb\endcsname%
      \put(176,1723){\rotatebox{-270}{\makebox(0,0){\strut{}}}}%
      \put(2282,-66){\makebox(0,0){\strut{}}}%
      \put(2282,3557){\makebox(0,0){\strut{}ASQ tensor ratios (fine)}}%
    }%
    \gplgaddtomacro\gplfronttext{%
    }%
    \gplbacktext
    \put(0,0){\includegraphics{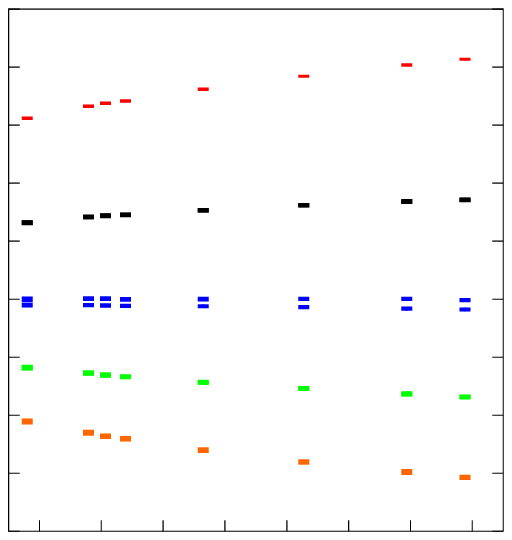}}%
    \gplfronttext
  \end{picture}%
\endgroup

%% file: figs/HYP_fine_ratios_scalars.tex
\begingroup
  \makeatletter
  \providecommand\color[2][]{%
    \GenericError{(gnuplot) \space\space\space\@spaces}{%
      Package color not loaded in conjunction with
      terminal option `colourtext'%
    }{See the gnuplot documentation for explanation.%
    }{Either use 'blacktext' in gnuplot or load the package
      color.sty in LaTeX.}%
    \renewcommand\color[2][]{}%
  }%
  \providecommand\includegraphics[2][]{%
    \GenericError{(gnuplot) \space\space\space\@spaces}{%
      Package graphicx or graphics not loaded%
    }{See the gnuplot documentation for explanation.%
    }{The gnuplot epslatex terminal needs graphicx.sty or graphics.sty.}%
    \renewcommand\includegraphics[2][]{}%
  }%
  \providecommand\rotatebox[2]{#2}%
  \@ifundefined{ifGPcolor}{%
    \newif\ifGPcolor
    \GPcolortrue
  }{}%
  \@ifundefined{ifGPblacktext}{%
    \newif\ifGPblacktext
    \GPblacktexttrue
  }{}%
  \let\gplgaddtomacro\g@addto@macro
  \gdef\gplbacktext{}%
  \gdef\gplfronttext{}%
  \makeatother
  \ifGPblacktext
    \def\colorrgb#1{}%
    \def\colorgray#1{}%
  \else
    \ifGPcolor
      \def\colorrgb#1{\color[rgb]{#1}}%
      \def\colorgray#1{\color[gray]{#1}}%
      \expandafter\def\csname LTw\endcsname{\color{white}}%
      \expandafter\def\csname LTb\endcsname{\color{black}}%
      \expandafter\def\csname LTa\endcsname{\color{black}}%
      \expandafter\def\csname LT0\endcsname{\color[rgb]{1,0,0}}%
      \expandafter\def\csname LT1\endcsname{\color[rgb]{0,1,0}}%
      \expandafter\def\csname LT2\endcsname{\color[rgb]{0,0,1}}%
      \expandafter\def\csname LT3\endcsname{\color[rgb]{1,0,1}}%
      \expandafter\def\csname LT4\endcsname{\color[rgb]{0,1,1}}%
      \expandafter\def\csname LT5\endcsname{\color[rgb]{1,1,0}}%
      \expandafter\def\csname LT6\endcsname{\color[rgb]{0,0,0}}%
      \expandafter\def\csname LT7\endcsname{\color[rgb]{1,0.3,0}}%
      \expandafter\def\csname LT8\endcsname{\color[rgb]{0.5,0.5,0.5}}%
    \else
      \def\colorrgb#1{\color{black}}%
      \def\colorgray#1{\color[gray]{#1}}%
      \expandafter\def\csname LTw\endcsname{\color{white}}%
      \expandafter\def\csname LTb\endcsname{\color{black}}%
      \expandafter\def\csname LTa\endcsname{\color{black}}%
      \expandafter\def\csname LT0\endcsname{\color{black}}%
      \expandafter\def\csname LT1\endcsname{\color{black}}%
      \expandafter\def\csname LT2\endcsname{\color{black}}%
      \expandafter\def\csname LT3\endcsname{\color{black}}%
      \expandafter\def\csname LT4\endcsname{\color{black}}%
      \expandafter\def\csname LT5\endcsname{\color{black}}%
      \expandafter\def\csname LT6\endcsname{\color{black}}%
      \expandafter\def\csname LT7\endcsname{\color{black}}%
      \expandafter\def\csname LT8\endcsname{\color{black}}%
    \fi
  \fi
  \setlength{\unitlength}{0.0500bp}%
  \begin{picture}(4320.00,4320.00)%
    \gplgaddtomacro\gplbacktext{%
      \colorrgb{0.00,0.00,0.00}%
      \put(946,704){\makebox(0,0)[r]{\strut{}1.00}}%
      \colorrgb{0.00,0.00,0.00}%
      \put(946,1032){\makebox(0,0)[r]{\strut{}1.05}}%
      \colorrgb{0.00,0.00,0.00}%
      \put(946,1361){\makebox(0,0)[r]{\strut{}1.10}}%
      \colorrgb{0.00,0.00,0.00}%
      \put(946,1689){\makebox(0,0)[r]{\strut{}1.15}}%
      \colorrgb{0.00,0.00,0.00}%
      \put(946,2017){\makebox(0,0)[r]{\strut{}1.20}}%
      \colorrgb{0.00,0.00,0.00}%
      \put(946,2346){\makebox(0,0)[r]{\strut{}1.25}}%
      \colorrgb{0.00,0.00,0.00}%
      \put(946,2674){\makebox(0,0)[r]{\strut{}1.30}}%
      \colorrgb{0.00,0.00,0.00}%
      \put(946,3002){\makebox(0,0)[r]{\strut{}1.35}}%
      \colorrgb{0.00,0.00,0.00}%
      \put(946,3331){\makebox(0,0)[r]{\strut{}1.40}}%
      \colorrgb{0.00,0.00,0.00}%
      \put(946,3659){\makebox(0,0)[r]{\strut{}1.45}}%
      \colorrgb{0.00,0.00,0.00}%
      \put(1256,484){\makebox(0,0){\strut{}0.6}}%
      \colorrgb{0.00,0.00,0.00}%
      \put(1611,484){\makebox(0,0){\strut{}0.8}}%
      \colorrgb{0.00,0.00,0.00}%
      \put(1967,484){\makebox(0,0){\strut{}1.0}}%
      \colorrgb{0.00,0.00,0.00}%
      \put(2323,484){\makebox(0,0){\strut{}1.2}}%
      \colorrgb{0.00,0.00,0.00}%
      \put(2678,484){\makebox(0,0){\strut{}1.4}}%
      \colorrgb{0.00,0.00,0.00}%
      \put(3034,484){\makebox(0,0){\strut{}1.6}}%
      \colorrgb{0.00,0.00,0.00}%
      \put(3390,484){\makebox(0,0){\strut{}1.8}}%
      \colorrgb{0.00,0.00,0.00}%
      \put(3745,484){\makebox(0,0){\strut{}2.0}}%
      \csname LTb\endcsname%
      \put(176,2181){\makebox(0,0){\Large $\frac{Z_{\mathbf{1} \otimes \xi_F}}            {Z_{\mathbf{1} \otimes \mathbf{1}}}$}}%
      \put(2500,154){\makebox(0,0){\strut{}$(ap)^2$}}%
      \put(2500,3989){\makebox(0,0){\strut{}HYP scalar ratios (fine)}}%
    }%
    \gplgaddtomacro\gplfronttext{%
    }%
    \gplbacktext
    \put(0,0){\includegraphics{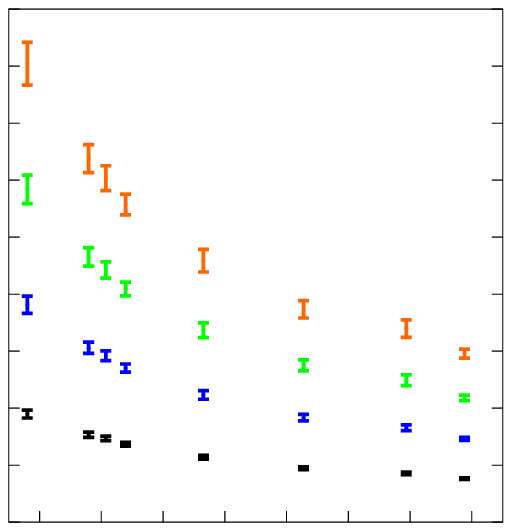}}%
    \gplfronttext
  \end{picture}%
\endgroup

%% file: figs/ASQ_fine_ratios_scalars.tex
\begingroup
  \makeatletter
  \providecommand\color[2][]{%
    \GenericError{(gnuplot) \space\space\space\@spaces}{%
      Package color not loaded in conjunction with
      terminal option `colourtext'%
    }{See the gnuplot documentation for explanation.%
    }{Either use 'blacktext' in gnuplot or load the package
      color.sty in LaTeX.}%
    \renewcommand\color[2][]{}%
  }%
  \providecommand\includegraphics[2][]{%
    \GenericError{(gnuplot) \space\space\space\@spaces}{%
      Package graphicx or graphics not loaded%
    }{See the gnuplot documentation for explanation.%
    }{The gnuplot epslatex terminal needs graphicx.sty or graphics.sty.}%
    \renewcommand\includegraphics[2][]{}%
  }%
  \providecommand\rotatebox[2]{#2}%
  \@ifundefined{ifGPcolor}{%
    \newif\ifGPcolor
    \GPcolortrue
  }{}%
  \@ifundefined{ifGPblacktext}{%
    \newif\ifGPblacktext
    \GPblacktexttrue
  }{}%
  \let\gplgaddtomacro\g@addto@macro
  \gdef\gplbacktext{}%
  \gdef\gplfronttext{}%
  \makeatother
  \ifGPblacktext
    \def\colorrgb#1{}%
    \def\colorgray#1{}%
  \else
    \ifGPcolor
      \def\colorrgb#1{\color[rgb]{#1}}%
      \def\colorgray#1{\color[gray]{#1}}%
      \expandafter\def\csname LTw\endcsname{\color{white}}%
      \expandafter\def\csname LTb\endcsname{\color{black}}%
      \expandafter\def\csname LTa\endcsname{\color{black}}%
      \expandafter\def\csname LT0\endcsname{\color[rgb]{1,0,0}}%
      \expandafter\def\csname LT1\endcsname{\color[rgb]{0,1,0}}%
      \expandafter\def\csname LT2\endcsname{\color[rgb]{0,0,1}}%
      \expandafter\def\csname LT3\endcsname{\color[rgb]{1,0,1}}%
      \expandafter\def\csname LT4\endcsname{\color[rgb]{0,1,1}}%
      \expandafter\def\csname LT5\endcsname{\color[rgb]{1,1,0}}%
      \expandafter\def\csname LT6\endcsname{\color[rgb]{0,0,0}}%
      \expandafter\def\csname LT7\endcsname{\color[rgb]{1,0.3,0}}%
      \expandafter\def\csname LT8\endcsname{\color[rgb]{0.5,0.5,0.5}}%
    \else
      \def\colorrgb#1{\color{black}}%
      \def\colorgray#1{\color[gray]{#1}}%
      \expandafter\def\csname LTw\endcsname{\color{white}}%
      \expandafter\def\csname LTb\endcsname{\color{black}}%
      \expandafter\def\csname LTa\endcsname{\color{black}}%
      \expandafter\def\csname LT0\endcsname{\color{black}}%
      \expandafter\def\csname LT1\endcsname{\color{black}}%
      \expandafter\def\csname LT2\endcsname{\color{black}}%
      \expandafter\def\csname LT3\endcsname{\color{black}}%
      \expandafter\def\csname LT4\endcsname{\color{black}}%
      \expandafter\def\csname LT5\endcsname{\color{black}}%
      \expandafter\def\csname LT6\endcsname{\color{black}}%
      \expandafter\def\csname LT7\endcsname{\color{black}}%
      \expandafter\def\csname LT8\endcsname{\color{black}}%
    \fi
  \fi
  \setlength{\unitlength}{0.0500bp}%
  \begin{picture}(4104.00,4320.00)%
    \gplgaddtomacro\gplbacktext{%
      \colorrgb{0.00,0.00,0.00}%
      \put(726,704){\makebox(0,0)[r]{\strut{}1.15}}%
      \colorrgb{0.00,0.00,0.00}%
      \put(726,1073){\makebox(0,0)[r]{\strut{}1.20}}%
      \colorrgb{0.00,0.00,0.00}%
      \put(726,1443){\makebox(0,0)[r]{\strut{}1.25}}%
      \colorrgb{0.00,0.00,0.00}%
      \put(726,1812){\makebox(0,0)[r]{\strut{}1.30}}%
      \colorrgb{0.00,0.00,0.00}%
      \put(726,2182){\makebox(0,0)[r]{\strut{}1.35}}%
      \colorrgb{0.00,0.00,0.00}%
      \put(726,2551){\makebox(0,0)[r]{\strut{}1.40}}%
      \colorrgb{0.00,0.00,0.00}%
      \put(726,2920){\makebox(0,0)[r]{\strut{}1.45}}%
      \colorrgb{0.00,0.00,0.00}%
      \put(726,3290){\makebox(0,0)[r]{\strut{}1.50}}%
      \colorrgb{0.00,0.00,0.00}%
      \put(726,3659){\makebox(0,0)[r]{\strut{}1.55}}%
      \colorrgb{0.00,0.00,0.00}%
      \put(1036,484){\makebox(0,0){\strut{}0.6}}%
      \colorrgb{0.00,0.00,0.00}%
      \put(1392,484){\makebox(0,0){\strut{}0.8}}%
      \colorrgb{0.00,0.00,0.00}%
      \put(1748,484){\makebox(0,0){\strut{}1.0}}%
      \colorrgb{0.00,0.00,0.00}%
      \put(2104,484){\makebox(0,0){\strut{}1.2}}%
      \colorrgb{0.00,0.00,0.00}%
      \put(2461,484){\makebox(0,0){\strut{}1.4}}%
      \colorrgb{0.00,0.00,0.00}%
      \put(2817,484){\makebox(0,0){\strut{}1.6}}%
      \colorrgb{0.00,0.00,0.00}%
      \put(3173,484){\makebox(0,0){\strut{}1.8}}%
      \colorrgb{0.00,0.00,0.00}%
      \put(3529,484){\makebox(0,0){\strut{}2.0}}%
      \csname LTb\endcsname%
      \put(176,2181){\rotatebox{-270}{\makebox(0,0){\strut{}}}}%
      \put(2282,154){\makebox(0,0){\strut{}$(ap)^2$}}%
      \put(2282,3989){\makebox(0,0){\strut{}ASQ scalar ratios (fine)}}%
    }%
    \gplgaddtomacro\gplfronttext{%
    }%
    \gplbacktext
    \put(0,0){\includegraphics{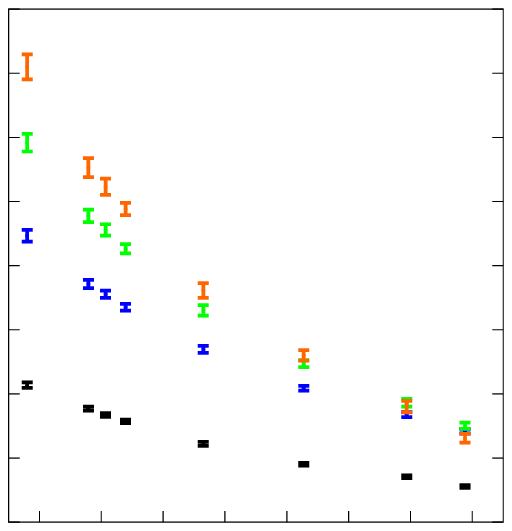}}%
    \gplfronttext
  \end{picture}%
\endgroup

%% file: figs/SminusP.tex
\begingroup
  \makeatletter
  \providecommand\color[2][]{%
    \GenericError{(gnuplot) \space\space\space\@spaces}{%
      Package color not loaded in conjunction with
      terminal option `colourtext'%
    }{See the gnuplot documentation for explanation.%
    }{Either use 'blacktext' in gnuplot or load the package
      color.sty in LaTeX.}%
    \renewcommand\color[2][]{}%
  }%
  \providecommand\includegraphics[2][]{%
    \GenericError{(gnuplot) \space\space\space\@spaces}{%
      Package graphicx or graphics not loaded%
    }{See the gnuplot documentation for explanation.%
    }{The gnuplot epslatex terminal needs graphicx.sty or graphics.sty.}%
    \renewcommand\includegraphics[2][]{}%
  }%
  \providecommand\rotatebox[2]{#2}%
  \@ifundefined{ifGPcolor}{%
    \newif\ifGPcolor
    \GPcolortrue
  }{}%
  \@ifundefined{ifGPblacktext}{%
    \newif\ifGPblacktext
    \GPblacktexttrue
  }{}%
  \let\gplgaddtomacro\g@addto@macro
  \gdef\gplbacktext{}%
  \gdef\gplfronttext{}%
  \makeatother
  \ifGPblacktext
    \def\colorrgb#1{}%
    \def\colorgray#1{}%
  \else
    \ifGPcolor
      \def\colorrgb#1{\color[rgb]{#1}}%
      \def\colorgray#1{\color[gray]{#1}}%
      \expandafter\def\csname LTw\endcsname{\color{white}}%
      \expandafter\def\csname LTb\endcsname{\color{black}}%
      \expandafter\def\csname LTa\endcsname{\color{black}}%
      \expandafter\def\csname LT0\endcsname{\color[rgb]{1,0,0}}%
      \expandafter\def\csname LT1\endcsname{\color[rgb]{0,1,0}}%
      \expandafter\def\csname LT2\endcsname{\color[rgb]{0,0,1}}%
      \expandafter\def\csname LT3\endcsname{\color[rgb]{1,0,1}}%
      \expandafter\def\csname LT4\endcsname{\color[rgb]{0,1,1}}%
      \expandafter\def\csname LT5\endcsname{\color[rgb]{1,1,0}}%
      \expandafter\def\csname LT6\endcsname{\color[rgb]{0,0,0}}%
      \expandafter\def\csname LT7\endcsname{\color[rgb]{1,0.3,0}}%
      \expandafter\def\csname LT8\endcsname{\color[rgb]{0.5,0.5,0.5}}%
    \else
      \def\colorrgb#1{\color{black}}%
      \def\colorgray#1{\color[gray]{#1}}%
      \expandafter\def\csname LTw\endcsname{\color{white}}%
      \expandafter\def\csname LTb\endcsname{\color{black}}%
      \expandafter\def\csname LTa\endcsname{\color{black}}%
      \expandafter\def\csname LT0\endcsname{\color{black}}%
      \expandafter\def\csname LT1\endcsname{\color{black}}%
      \expandafter\def\csname LT2\endcsname{\color{black}}%
      \expandafter\def\csname LT3\endcsname{\color{black}}%
      \expandafter\def\csname LT4\endcsname{\color{black}}%
      \expandafter\def\csname LT5\endcsname{\color{black}}%
      \expandafter\def\csname LT6\endcsname{\color{black}}%
      \expandafter\def\csname LT7\endcsname{\color{black}}%
      \expandafter\def\csname LT8\endcsname{\color{black}}%
    \fi
  \fi
  \setlength{\unitlength}{0.0500bp}%
  \begin{picture}(4320.00,4320.00)%
    \gplgaddtomacro\gplbacktext{%
      \colorrgb{0.00,0.00,0.00}%
      \put(814,704){\makebox(0,0)[r]{\strut{}0.0}}%
      \colorrgb{0.00,0.00,0.00}%
      \put(814,1417){\makebox(0,0)[r]{\strut{}0.1}}%
      \colorrgb{0.00,0.00,0.00}%
      \put(814,2130){\makebox(0,0)[r]{\strut{}0.2}}%
      \colorrgb{0.00,0.00,0.00}%
      \put(814,2843){\makebox(0,0)[r]{\strut{}0.3}}%
      \colorrgb{0.00,0.00,0.00}%
      \put(814,3556){\makebox(0,0)[r]{\strut{}0.4}}%
      \colorrgb{0.00,0.00,0.00}%
      \put(946,484){\makebox(0,0){\strut{} 0.5}}%
      \colorrgb{0.00,0.00,0.00}%
      \put(1371,484){\makebox(0,0){\strut{} 1}}%
      \colorrgb{0.00,0.00,0.00}%
      \put(1797,484){\makebox(0,0){\strut{} 1.5}}%
      \colorrgb{0.00,0.00,0.00}%
      \put(2222,484){\makebox(0,0){\strut{} 2}}%
      \colorrgb{0.00,0.00,0.00}%
      \put(2647,484){\makebox(0,0){\strut{} 2.5}}%
      \colorrgb{0.00,0.00,0.00}%
      \put(3072,484){\makebox(0,0){\strut{} 3}}%
      \colorrgb{0.00,0.00,0.00}%
      \put(3498,484){\makebox(0,0){\strut{} 3.5}}%
      \colorrgb{0.00,0.00,0.00}%
      \put(3923,484){\makebox(0,0){\strut{} 4}}%
      \csname LTb\endcsname%
      \put(176,2379){\makebox(0,0){\Large $\frac{\Lambda_S - \Lambda_P}{\Lambda_{\text{ave}}}$}}%
      \put(2434,154){\makebox(0,0){\strut{}$(ap)^2$}}%
    }%
    \gplgaddtomacro\gplfronttext{%
      \csname LTb\endcsname%
      \put(3137,3882){\makebox(0,0)[r]{\strut{}E}}%
      \csname LTb\endcsname%
      \put(3137,3662){\makebox(0,0)[r]{\strut{}NE}}%
    }%
    \gplbacktext
    \put(0,0){\includegraphics{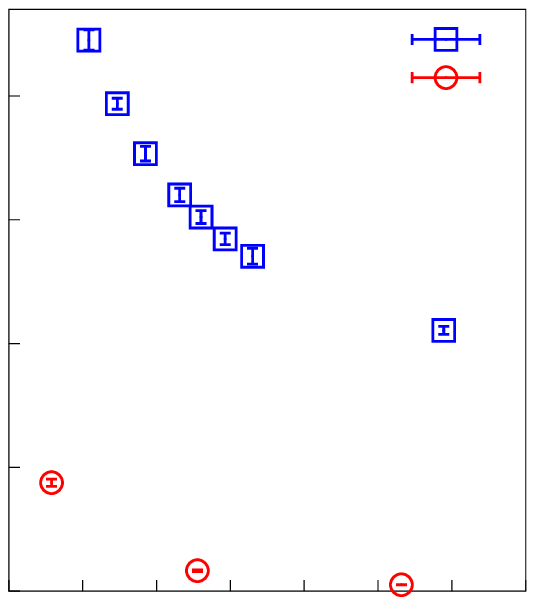}}%
    \gplfronttext
  \end{picture}%
\endgroup

%% file: figs/LambdaS.tex
\begingroup
  \makeatletter
  \providecommand\color[2][]{%
    \GenericError{(gnuplot) \space\space\space\@spaces}{%
      Package color not loaded in conjunction with
      terminal option `colourtext'%
    }{See the gnuplot documentation for explanation.%
    }{Either use 'blacktext' in gnuplot or load the package
      color.sty in LaTeX.}%
    \renewcommand\color[2][]{}%
  }%
  \providecommand\includegraphics[2][]{%
    \GenericError{(gnuplot) \space\space\space\@spaces}{%
      Package graphicx or graphics not loaded%
    }{See the gnuplot documentation for explanation.%
    }{The gnuplot epslatex terminal needs graphicx.sty or graphics.sty.}%
    \renewcommand\includegraphics[2][]{}%
  }%
  \providecommand\rotatebox[2]{#2}%
  \@ifundefined{ifGPcolor}{%
    \newif\ifGPcolor
    \GPcolortrue
  }{}%
  \@ifundefined{ifGPblacktext}{%
    \newif\ifGPblacktext
    \GPblacktexttrue
  }{}%
  \let\gplgaddtomacro\g@addto@macro
  \gdef\gplbacktext{}%
  \gdef\gplfronttext{}%
  \makeatother
  \ifGPblacktext
    \def\colorrgb#1{}%
    \def\colorgray#1{}%
  \else
    \ifGPcolor
      \def\colorrgb#1{\color[rgb]{#1}}%
      \def\colorgray#1{\color[gray]{#1}}%
      \expandafter\def\csname LTw\endcsname{\color{white}}%
      \expandafter\def\csname LTb\endcsname{\color{black}}%
      \expandafter\def\csname LTa\endcsname{\color{black}}%
      \expandafter\def\csname LT0\endcsname{\color[rgb]{1,0,0}}%
      \expandafter\def\csname LT1\endcsname{\color[rgb]{0,1,0}}%
      \expandafter\def\csname LT2\endcsname{\color[rgb]{0,0,1}}%
      \expandafter\def\csname LT3\endcsname{\color[rgb]{1,0,1}}%
      \expandafter\def\csname LT4\endcsname{\color[rgb]{0,1,1}}%
      \expandafter\def\csname LT5\endcsname{\color[rgb]{1,1,0}}%
      \expandafter\def\csname LT6\endcsname{\color[rgb]{0,0,0}}%
      \expandafter\def\csname LT7\endcsname{\color[rgb]{1,0.3,0}}%
      \expandafter\def\csname LT8\endcsname{\color[rgb]{0.5,0.5,0.5}}%
    \else
      \def\colorrgb#1{\color{black}}%
      \def\colorgray#1{\color[gray]{#1}}%
      \expandafter\def\csname LTw\endcsname{\color{white}}%
      \expandafter\def\csname LTb\endcsname{\color{black}}%
      \expandafter\def\csname LTa\endcsname{\color{black}}%
      \expandafter\def\csname LT0\endcsname{\color{black}}%
      \expandafter\def\csname LT1\endcsname{\color{black}}%
      \expandafter\def\csname LT2\endcsname{\color{black}}%
      \expandafter\def\csname LT3\endcsname{\color{black}}%
      \expandafter\def\csname LT4\endcsname{\color{black}}%
      \expandafter\def\csname LT5\endcsname{\color{black}}%
      \expandafter\def\csname LT6\endcsname{\color{black}}%
      \expandafter\def\csname LT7\endcsname{\color{black}}%
      \expandafter\def\csname LT8\endcsname{\color{black}}%
    \fi
  \fi
  \setlength{\unitlength}{0.0500bp}%
  \begin{picture}(4320.00,4320.00)%
    \gplgaddtomacro\gplbacktext{%
      \colorrgb{0.00,0.00,0.00}%
      \put(814,983){\makebox(0,0)[r]{\strut{}1.1}}%
      \colorrgb{0.00,0.00,0.00}%
      \put(814,1542){\makebox(0,0)[r]{\strut{}1.3}}%
      \colorrgb{0.00,0.00,0.00}%
      \put(814,2100){\makebox(0,0)[r]{\strut{}1.5}}%
      \colorrgb{0.00,0.00,0.00}%
      \put(814,2659){\makebox(0,0)[r]{\strut{}1.7}}%
      \colorrgb{0.00,0.00,0.00}%
      \put(814,3217){\makebox(0,0)[r]{\strut{}1.9}}%
      \colorrgb{0.00,0.00,0.00}%
      \put(814,3776){\makebox(0,0)[r]{\strut{}2.1}}%
      \colorrgb{0.00,0.00,0.00}%
      \put(1244,484){\makebox(0,0){\strut{} 0.01}}%
      \colorrgb{0.00,0.00,0.00}%
      \put(2435,484){\makebox(0,0){\strut{} 0.02}}%
      \colorrgb{0.00,0.00,0.00}%
      \put(3625,484){\makebox(0,0){\strut{} 0.03}}%
      \csname LTb\endcsname%
      \put(176,2379){\makebox(0,0){\strut{}$\quad \Lambda_S$}}%
      \put(2434,154){\makebox(0,0){\strut{}$am$}}%
    }%
    \gplgaddtomacro\gplfronttext{%
      \csname LTb\endcsname%
      \put(3137,3882){\makebox(0,0)[r]{\strut{}E}}%
      \csname LTb\endcsname%
      \put(3137,3662){\makebox(0,0)[r]{\strut{}NE}}%
    }%
    \gplbacktext
    \put(0,0){\includegraphics{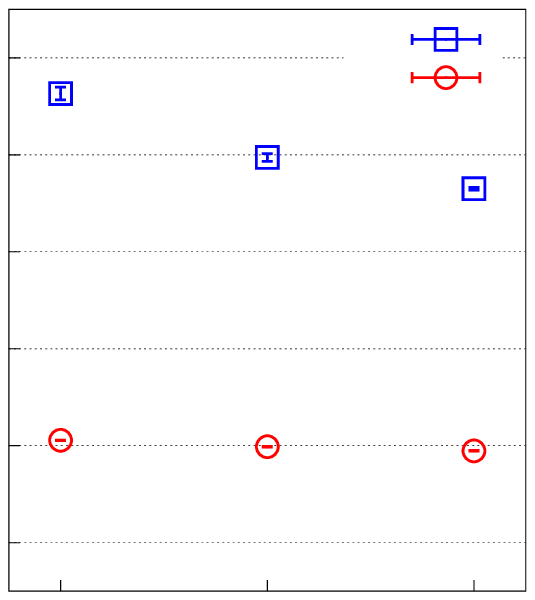}}%
    \gplfronttext
  \end{picture}%
\endgroup